\documentclass[a4,11pt,twoside]{article}

\usepackage{asp2010}

\resetcounters

\begin{document}
\title{Analysing ALMA data with CASA}

\author{Dirk Petry for the CASA development team$^*$}

\affil{ALMA Regional Centre, ESO, Karl-Schwarzschild-Str. 2, D-85748 Garching}
\affil{$^*$ see acknowledgements}

\begin{abstract}
The radio astronomical data analysis package CASA was selected to be the 
designated tool for observers to analyse the data from the Atacama Large mm/sub-mm Array (ALMA) 
which is under construction and has recently started taking its first science data (Cycle 0). 
CASA is a large package which is being developed by NRAO with major 
contributions from ESO and NAOJ.
Generally, all radio data from interferometers and single dish observatories can be 
analysed with CASA, but the development focuses presently on the needs of the new observatories 
EVLA and ALMA. This article describes the main features of CASA and the typical analysis steps
for ALMA data.
\end{abstract}

\vspace{-5mm}
\section{Introduction}
The Atacama Large Millimeter/submillimeter Array (ALMA)
is a major new facility for world astronomy. When completed 
in 2013, ALMA will consist of a giant array of 12-m antennas, 
with baselines up to 16 km, and an additional compact array 
of 7-m and 12-m antennas to greatly enhance ALMA's ability 
to image extended targets. ALMA in Cycle 0 is outfitted 
with state-of-the-art receivers that cover atmospheric windows 
from 84~GHz to 720~GHz (3~mm to 0.42~mm). Construction of ALMA 
started in 2003 and will be completed in 2013. Science 
observations have started in 2011 with 16 antennas and four 
receiver bands ({\it Early Science}). The ALMA project is an 
international collaboration between Europe, East Asia and 
North America in cooperation with the Republic of Chile.
The official project website for scientists is the {\it Science Portal}
\url{http://www.almascience.org}. Proposal cycle 0 started
with a call on 31 March 2011. 

The average daily science data volume generated by ALMA is expected to be roughly one TByte.  
All ALMA data is stored in the
{\it ALMA Science Data Model} (ASDM) format \citep{viallefond_2006} which
in its present implementation is a collection of XML and binary MIME
encoded files.

The final offline data reduction of the archive ASDM data needs a software package 
capable of handling all features of the ALMA data including the high spectral resolution and the
large data volume. In 2003, the {\it CASA} package was selected 
by the project as the ALMA data reduction software.
 
\section{The CASA data analysis package}
CASA (Common Astronomy Software Applications) is a general package for all
radio-astronomical data analysis. The development team
is led by NRAO with major contributions from ESO and NAOJ. Documentation is
available from the CASA homepage \url{http://casa.nrao.edu}. It includes a comprehensive
manual \citep{ott_2011}. 
Since 2003, the development focuses on
delivering the functionality necessary for the analysis of data from the
Expanded Very Large Array (EVLA) and ALMA.

\subsection{Technical overview}
CASA is a modern software package which was designed for easy maintenance,
high performance, and user friendliness. 
The underlying set of libraries which implements the fundamental functionality used for
data handling, transformations, calibration, and imaging, is {\it casacore}
(\url{http://code.google.com/p/casacore}). casacore is mostly written in C++. 
The actual CASA package includes a version
of casacore and extends it with
(a) observatory-specific C++ libraries for data import and export,
(b) C++ libraries implementing higher level calibration and imaging algorithms,
(c) C++ libraries for image and spectral viewing and analysis,
(d) C++ code for graphical user interface (GUI) applications for data visualisation and editing,
(e) C++ code and XML definitions for the binding of CASA to the Python scripting language,
(f) Python code for convenience scripts implementing common analysis procedures (``tasks''),
(g) reference tables of astronomical data, spectroscopic data, and observatory parameters.
Furthermore, the ASAP package for single-dish (SD) spectral analysis is included.
CASA developers also contribute to casacore and ASAP.

The supported computer platforms of CASA are the common Linux distributions (32 and 64 bit) 
and Mac OSX 10.6 and 10.7.
CASA comes as an easy to install binary distribution in a tar file (Linux) or DMG file (Mac OS) 
containing (depending on the platform) all necessary libraries from 
external packages (such as blas, boost, dbus, fftw, lapack, Python, xerces, wcslib)
to simplify installation and avoid version conflicts.
The latest release as of November 2011 is CASA 3.3.0. The binary release and the source code are available under GNU Public License via
\url{http://casa.nrao.edu}.

\subsection{User interface}
The user interacts with CASA via the commandline in the {\it casapy} shell which
uses iPython (interactive Python) to achieve a MATLAB-like work environment.
In addition, CASA offers several stand-alone GUI applications like the {\it viewer} and {\it plotms}
for data visualisation and editing (using the Qt widget set).
The command line interface has two levels:
\begin{description}
\item[Tools:] Essentially all functionality of the CASA libraries is accessible via a set of
   Python objects, the {\it tools}. The methods of these tools have detailed documentation which
   can be accessed in the standard Pyhton way through the help command.
\item[Tasks:] Selected common analysis procedures (which could in principle already be performed
   by using the tools) are available as parametrised built-in Python scripts. They are called the
   CASA {\it tasks}. These scripts are designed to be robust and user friendly. They are augmented 
   by a system of shared global parameters which can be
   stored and recalled separately for each task and automatically checked for correctness before
   task execution.   
   Users can extend the set of tasks with their own scripts. 
\end{description}

\section{ALMA data analysis}
Detailed annotated ALMA data analysis examples can be found among the CASA guides which
have been compiled by the ALMA CASA subsystem scientist C. Brogan together with 
experts from the three ALMA Regional Centres (ARCs) and the Joint ALMA Observatory (JAO). 
See \url{http://casaguides.nrao.edu}. In the following we discuss the typical
steps of an ALMA data analysis in general.

\subsection{Import}
The first step of any data analysis in CASA is the conversion of the input data into
the native CASA data format, the {\it Measurement Set} (MS) which is the casacore implementation
of the data model proposed by \citet{hamakeretal_1996}. The translation from the ASDM to the MS format
is achieved using the task {\it importasdm} which offers several options for controlling which subset
of the data is imported an how it is stored. 
For various technical reasons it is likely that the conversion to MS for ALMA Cycle 0 data will already
take place at the observatory and the user will receive data in MS format.

\subsection{Inspection and ``flagging''}
Once the data is available as an MS, the user can inspect them using tasks like {\it listobs} and {\it plotms}.
Consulting observing logs and looking at diagnostic plots of visibility amplitude and phase vs. time or frequency,
the user identifies patches of data which are not useful due to technical problems or bad weather conditions.
(As the data will have undergone a quality assurance procedure, there will be very little of such data.)
The corresponding data are then ``flagged'', i.e. marked as not to be used, either using point-and-click methods
in the GUIs or tasks like {\it flagdata}. Different states of the flagging information can be recorded and 
restored using the {\it flagmanager} task.

\subsection{Calibration}
While the calibration of single-dish data in CASA is essentially carried out in a single step using
the task {\it sdaverage}, the calibration of interferometric data is more involved and can require
several iterations depending on data quality and the strategy used for the calibrator observations.
It begins for ALMA data with the correction for
atmospheric phase (based on the Water Vapour Radiometer (WVR) measurements 
from each antenna), atmospheric emission/absorption (system temperature, ``$T_{\mathrm sys}$''), 
and possibly instrumental errors. All of these will be determined at the observatory and
the user will receive the necessary calibration tables and/or the calibrated MS.

The remaining interferometric calibration is then based on the tasks {\it gaincal} and {\it bandpass}. 
They permit the separate or combined fitting of calibration
solutions for visibility phase and amplitude to calibrator observations.   
The solutions are scaled to an absolute flux scale (based on a flux calibrator observation
with corresponding model) using the tasks {\it setjy} and {\it fluxscale}. 
All solutions are stored in so-called calibration tables which can be inspected using tasks like {\it plotcal}.
They are applied to the target data using the task {\it applycal}. 

\subsection{Imaging}
CASA offers implementations of all common and many new and experimental imaging algorithms for interferometic data
as a toolkit such that users can construct the sequence of analysis methods most adequate for their data.
A dedicated team of imaging experts works at NRAO to continuously extend
this kit. Presently, the access to all of the kit is through the task {\it clean}.
When studying spectral lines, the task {\it uvcontsub} serves to determine and subtract 
only weakly frequency-dependent spectral components (the continuum).
Single-dish imaging is performed using the task {\it sdimaging}.  

\subsection{Selfcalibration}
Experience with ALMA test data has shown that often so-called selfcalibration (selfcal)
is possible and leads to reliable improvements in the signal-to-noise ratio of the image.
Selfcal is possible if the target is bright and has image components which have a known
morphology. With a large number of baselines as in the case of ALMA,
the calibration solutions are overdetermined and the assumption of a known morphology can be used
to improve their quality. 
Selfcal is achieved in CASA via a combination of the calibration and imaging tasks (see above).
It is typically performed on a continuum image and then transferred to the entire dataset. 

\subsection{Image analysis}
The last step of the analysis which can be carried out in CASA is the statistical and morphological analysis
of the obtained images and spectra.
With tasks like {\it imfit}, {\it imstat}, {\it immoments}, and {\it specfit}, the user can
fit geometrical shapes to image and spectral features and calculate the common statistical parameters.
The image data are always also directly available for analysis in Python scripts written by the user.  
For the identification of spectral lines, a spectral line catalog and tasks like {\it slsearch} are included.  
The {\it viewer} permits the generation of publication-ready colour plots of images and spectra.  

\section{Prospects for the next releases}
The CASA release cycle foresees new releases every six months. Development for release 3.4 will
include performance improvements and parallelisation support, full polarisation calibration,
plotting and visualisation improvements. 

\acknowledgements
{\it CASA} is developed at NRAO and under NRAO management with
major contributions from ESO and NAOJ. 
Manager: J. Kern (NRAO), project scientist: J. Ott (NRAO), 
ALMA CASA subsystem scientist: C. Brogan (NRAO), developers at NRAO: 
S. Bhatnagar, K. Golap, R. Indebetouw,  J. Jacobs, D. Mehringer, 
G. Moellenbrock, S. Rankin, U. Rau, R. Reid, D. Schiebel, 
T. Tsutsumi, H. Ye, W. Young, developers at ESO: M. Caillat, 
S. Castro, J. Gonzales, M. Kuemmel,  D. Petry, developers at NAOJ: 
S. Kawakami, W. Kawasaki, K. Sugimoto, T. Nakazato,  
ALMA Regional Center CASA leads: D. Petry, M. Zwaan (ESO), 
D. Espada (NAOJ)

\bibliography{P114}

\begin{thebibliography}{}
\expandafter\ifx\csname natexlab\endcsname\relax\def\natexlab#1{#1}\fi
\expandafter\ifx\csname url\endcsname\relax
  \def\url#1{\texttt{#1}}\fi
\expandafter\ifx\csname urlprefix\endcsname\relax\def\urlprefix{URL }\fi
\providecommand{\eprint}[2][]{\url{#2}}

\bibitem[{Hamaker et~al.(1996)Hamaker, Bregman, \& Sault}]{hamakeretal_1996}
Hamaker, J., Bregman, J., \& Sault, R. 1996, A\&AS, 117, 137

\bibitem[{Ott(2011)}]{ott_2011}
Ott, J. 2011, CASA Synthesis \& Single Dish Reduction Reference Manual \&
  Cookbook (NRAO, version October 2011)

\bibitem[{Viallefond(2006)}]{viallefond_2006}
Viallefond, F. 2006, in ADASS XV, edited by C.~Gabriel, C.~Arviset, D.~Ponz, \&
  E.~Solano (San Francisco: ASP), vol. 351 of ASP Conf. Ser., 627

\end{thebibliography}

\end{document}